\documentclass[a4paper,12pt,aps]{revtex4}
\usepackage{amsmath,amssymb,lmodern}
\usepackage{bbold}
\usepackage{bm}
\usepackage{relsize}
\usepackage{braket}
\usepackage{bigints}

\usepackage{graphicx}

\newcommand{\average}[1]{\langle{#1}\rangle}

\DeclareMathOperator{\Tr}{Tr}

\renewcommand{\tt}{\textit}

\begin{document}

\title{Thermal behavior and entanglement in Pb-Pb and p-p
  collisions}
\author{X. Feal}
\author{C. Pajares}
\author{R.A. Vazquez}
\affiliation{ 
Instituto Galego de F\'{\i}sica de Altas Enerx\'{\i}as \& \\
Departamento de F\'{\i}sica de Part\'{\i}culas \\
Universidade de Santiago de Compostela, 15782 Santiago, SPAIN
}

\date{\today}
\begin{abstract}
The thermalization of the particles produced in collisions of small size
objects can be achieved by quantum entanglement of the partons of the initial
state as it was analyzed recently in proton-proton collisions. We extend such
study to Pb-Pb collisions and to different multiplicities of proton-proton
collisions. We observe that, in all cases, the effective temperature is
approximately proportional to the hard scale of the collision. We show that
such relation between the thermalization temperature and the hard scale can be
explained as a consequence of the clustering of the color sources. The
fluctuations on the number of parton states decreases with multiplicity in
Pb-Pb collisions as far as the width of the transverse momentum distributions
decreases, contrary to the p-p case. We relate these fluctuations to the
temperature time fluctuations by means of a Langevin equation for the white
noise due to the quench of a hard parton collision. 
\end{abstract}
\maketitle

\section{Introduction}

The presence of an exponential shape in the transverse momentum distribution
(TMD) of the produced particles in collisions of small size objects together
with the approximate thermal abundances of the hadron yields constitutes an
indicative sign of thermalization. This thermalization, however, can not be
achieved under the usual mechanism, namely final state interactions in the
form of several secondary collisions.

The emergence of this phenomenon has been recently studied
\cite{baker2017,kharzeev2017,bergers2017,bergers2018}, showing that
thermalization can be obtained during the rapid quench induced by the
collision due to the high degree of entanglement inside the partonic wave
functions of the colliding protons. Thus, the effective temperature obtained
from the TMD of the particles produced in the collision depends on the
momentum transfer, that is, it constitutes an ultraviolet cut-off on the
quantum modes resolved by the collision.  In diffractive processes with a
rapidity gap, the entire wave function of the proton is involved and no
entanglement entropy arises. Consequently, we expect no thermal radiation as
it has been observed.

In this paper we further explore the relation between parton entanglement and
thermalization by studying p-p and Pb-Pb collisions at different
multiplicities. In the second case we expect an interplay between
thermalization and final state interactions leading to some differences with
p-p collisions concerning the entanglement and thermalization.

We show that the TMD of both collisions at different multiplicities can be
fitted by the sum of an exponential plus a power like function, characterized
by a thermal like temperature $T_{\rm th}$ and a temperature scale 
$T_{\rm h}$ respectively. For any fixed multiplicity and in all collisions
the relation $4 T_{\rm th} \approx T_{\rm h}$ is satisfied. The power index $n$
describing the hard spectrum behaves differently in p-p and in Pb-Pb
collisions, showing the different behavior of the transverse momentum
fluctuations. This behavior and the relation between  $T_{\rm th}$ and
$T_{\rm h}$ can be naturally explained in the clustering of color sources.
The cluster size distribution of the clusters of overlapping strings found in
the collision coincides with the distribution of temperatures obtained as
solution of the Fokker-Planck equation associated to the linear Langevin
equation for a white Gaussian noise produced by a fast quench in a hard parton
collision. 

The organization of the paper is as follows. In section
\ref{entanglement_thermalization} we introduce the entanglement of the
partonic state following reference \cite{baker2017} and we analyze the TMD of
p-p and Pb-Pb collisions at different multiplicities. In section
\ref{discussion} we discuss the obtained results remarking the similarities
and differences of p-p and Pb-Pb collisions in connection with thermalization
and entanglement. We briefly discuss the clustering of color sources in
connection with the TMD in section \ref{clustering} and in the section
\ref{sec:langevin} we introduce the Langevin and Fokker-Planck equations to
study the time temperature fluctuations. Finally in section \ref{conclusions}
the conclusions are presented.

\section{Entanglement, thermalization, and transverse momentum distributions}
\label{entanglement_thermalization}

A hard process with momentum transfer $Q$ probes only the region of space $H$
of transverse size $1/Q$. Let us denote by $S$ the region of space
complementary of $H$. The proton is described by the wave function
\begin{equation}
|\Psi_{HS}\rangle = \sum_{n} \alpha_n |\Psi_{n}^{H}\rangle|\Psi_{n}^{S}\rangle,
\label{wavefunction}
\end{equation}
of a suitably chosen orthonormal sets of states $|\Psi_{n}^{H}\rangle$ and 
$|\Psi_{n}^{S}\rangle$ localized in the domains $H$ and $S$. In the parton
model this full orthonormal set of states is given by the Fock states with
different number $n$ of partons. The state (\ref{wavefunction}) can not be
separated into a product $|\varphi^{H}\rangle \otimes |\varphi^{S}\rangle$ and
therefore $|\Psi_{HS}\rangle$ is entangled. The density matrix of the mixed
state probed in region $H$ is
\begin{equation}
S_H= \Tr_S \rho_{SH}=\sum_n \langle
\Psi_n^S|\Psi_{HS}\rangle\langle\Psi_{HS}|\Psi_n^{S}\rangle = \sum_n
|\alpha_n|^2|\Psi_n^{H}\rangle\langle\Psi_n^{H}|,
\end{equation}
where $|\alpha_n|^2=p_n$ is the probability of having a state with $n$
partons, independently of whether their interaction is hard or soft. The Von
Neumann entropy of this state is given by
\begin{equation}
S=-\sum_n p_n\log p_n.
\label{VN_Entropy}
\end{equation}
We can consider that a high momentum partonic configuration of the proton when
the collision takes place undergoes a rapid quench due to the QCD
interaction. The onset $\tau$ of this hard interaction is given by the
hardness scale $Q$, $\tau\sim 1/Q$. Since $\tau$ is small the quench creates a
highly excited multi-particle state. The produced particles have thermal-like
exponential spectrum with an effective temperature $T\approx
(2\pi\tau)^{-1}\approx Q/2\pi$. Thus, the thermal spectrum can be originated
due to the event horizon formed by the acceleration of the color field \cite{kharzeev2005}-\cite{kharzeev2007}.
On the other hand, the comparison with LHC
data on hadron multiplicity distributions \cite{kharzeev2017} indicates that
the produced Boltzmann entropy is close to the entanglement entropy of
equation (\ref{VN_Entropy}).  

In reference \cite{baker2017}, the thermal component of charged hadron
transverse momentum distribution in p-p collisions at $\sqrt{s}=13$ TeV is
parameterized as \cite{bylinkin2014a,bylinkin2016,bylinkin2014b}
\begin{equation}
\frac{1}{N_{\rm ev}} \frac{1}{2\pi p_t} \frac{d^2N_{\rm ev}}{d\eta dp_t} =
A_{\rm th}\exp\big(-m_t/T_{\rm th}\big),
\label{thermal_parametrization}
\end{equation}
where $T_{\rm th}$ is the effective temperature and $m_t=\sqrt{m^2+p_t^2}$ is the
transverse mass. The hard scattering, meanwhile, is parameterized as,
\begin{equation}
\frac{1}{N_{\rm ev}}\frac{1}{2\pi p_t}\frac{d^2N_{\rm ev}}{d\eta d
  p_t}=A_{\rm h}\frac{1}{\bigg(1+\frac{m_t^2}{n T_{\rm h}^2}\bigg)^n},
\label{hard_parametrization}
\end{equation}
where the temperature $T_{\rm h}$ and the index $n$ are parameters determined
from the fit to the experimental data. The value $T_{\rm th}=0.17$ GeV was found
\cite{baker2017}, agreeing with the one expected from the extrapolation of the
relation
\begin{equation}
T_{\rm th} = 0.098 \bigg(\sqrt{\frac{s}{s_0}}\bigg)^{0.06} \; \;{\rm (GeV)},
\label{thermal_temperature}
\end{equation}
obtained at lower energies. Similarly the hard scale $T_h$ is given by the
relation
\begin{equation}
T_{\rm h} = 0.409 \bigg(\sqrt{\frac{s}{s_0}}\bigg)^{0.06 } \; \;{\rm (GeV)}.
\label{hard_temperature}
\end{equation}
At $\sqrt{s}=13$ TeV, the values found for the hard scale are $T_{\rm
  h}=0.72$GeV and $n=3.1$. We notice that from Equations
(\ref{thermal_temperature}) and (\ref{hard_temperature}) one finds
\begin{equation}
\frac{T_{\rm h}}{T_{\rm th}}\approx 4.2,
\label{thermalVshard}
\end{equation}
independently of the energy. The ratio of the particular values obtained in the
fit \cite{baker2017} are close to this values.

\begin{figure}[h]
\includegraphics[scale=1]{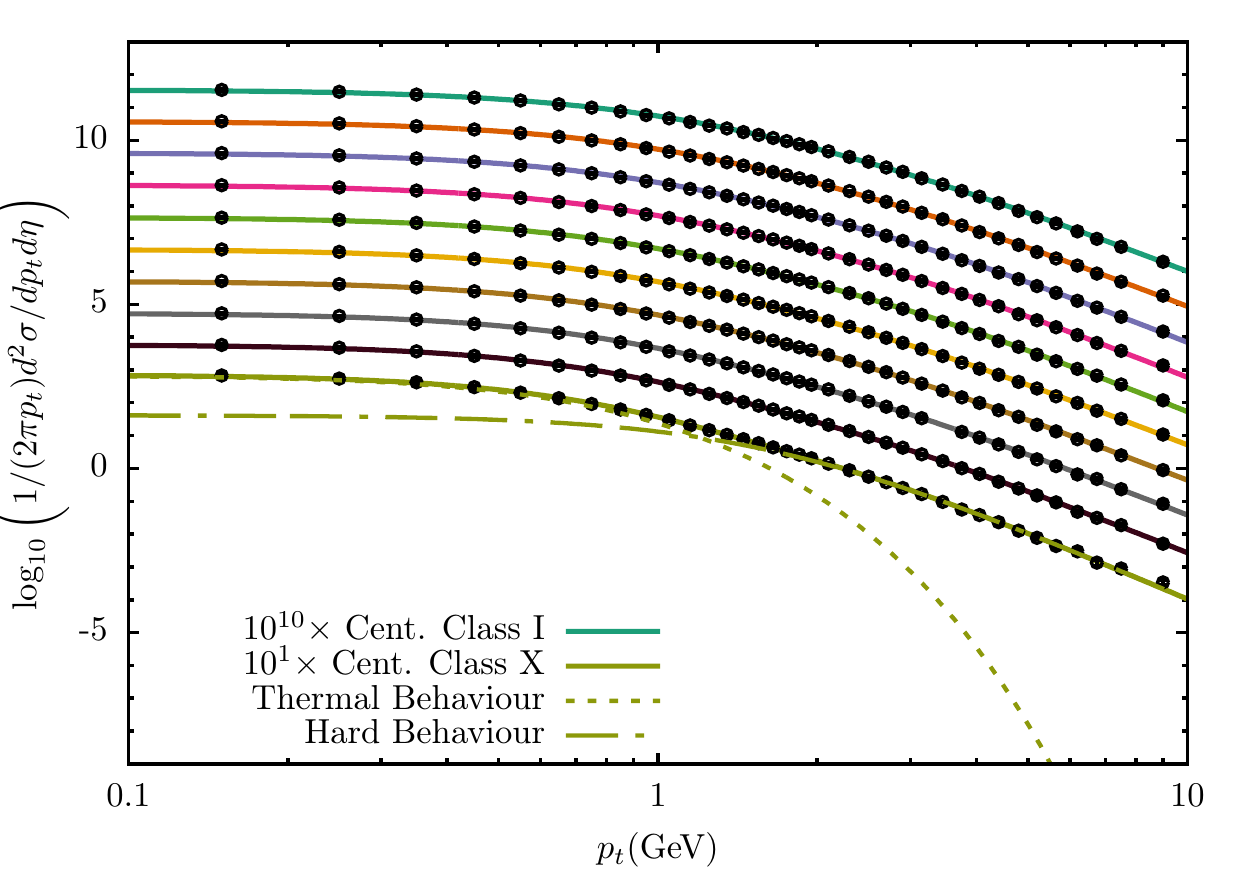}
\caption{Normalized differential $K_S^0$ production in p-p collisions at
  $\sqrt{s_{NN}}=$7 TeV as a function of transverse momentum for different
  classes of centralities. Centrality classes from I to X, in decreasing
  magnitude, correspond to charged particle productions of $dN_{\rm ch}/d\eta$ =
  21.3, 16.5, 13.5, 11.5, 10.1, 8.45, 6.72, 5.40, 3.90 and 2.26. Also shown
  are the thermal and hard components of the lowest centrality fit, by short
  dashed and long dashed lines, respectively.}
\label{fig:figure1}
\end{figure}

\begin{figure}[h]
\includegraphics[scale=1]{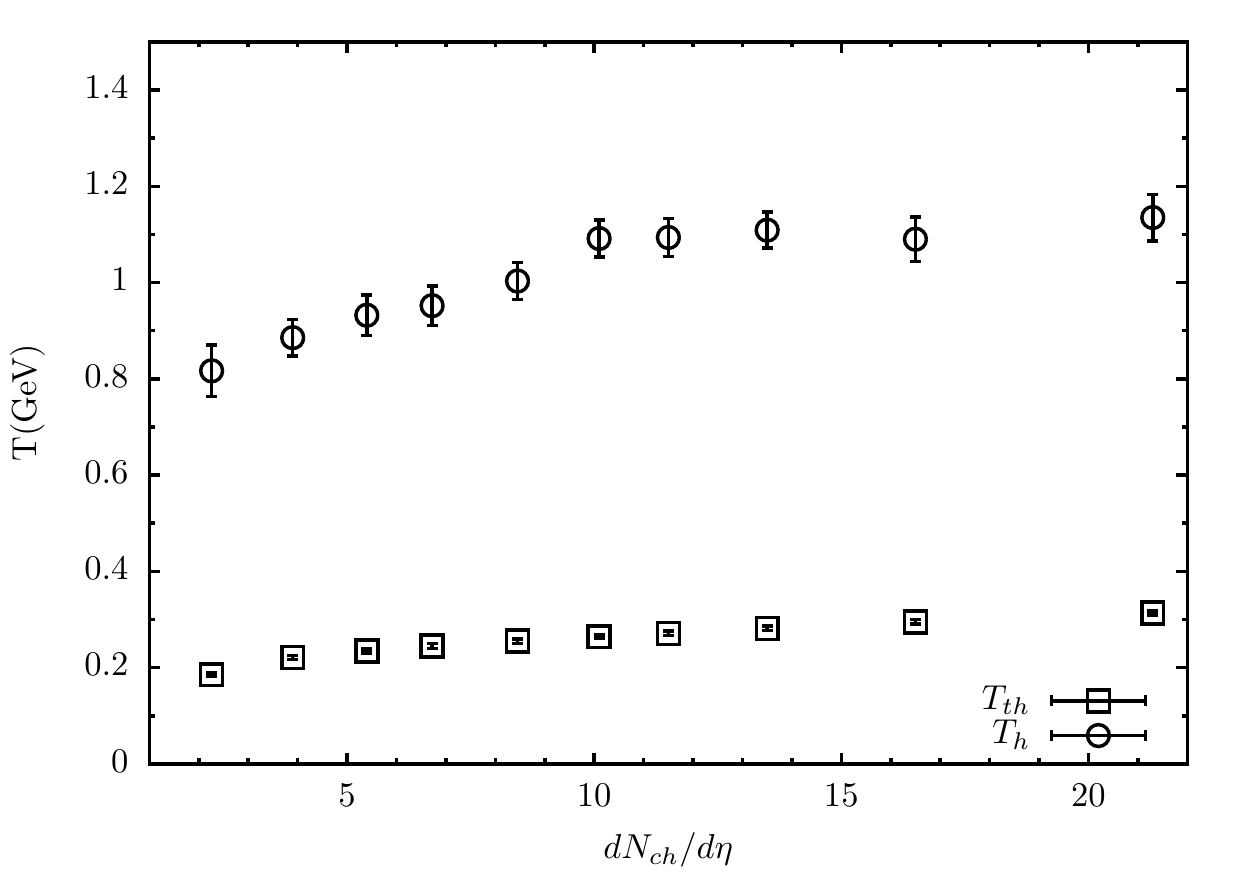}
\caption{ $T_{\rm th}$ and $T_{\rm h}$ as a function of centrality for
  $K_S^0$ production in p-p collisions at $\sqrt{s_{NN}}=$7 TeV. }
\label{fig:figure2}
\end{figure}

\begin{figure}[h]
\includegraphics[scale=1]{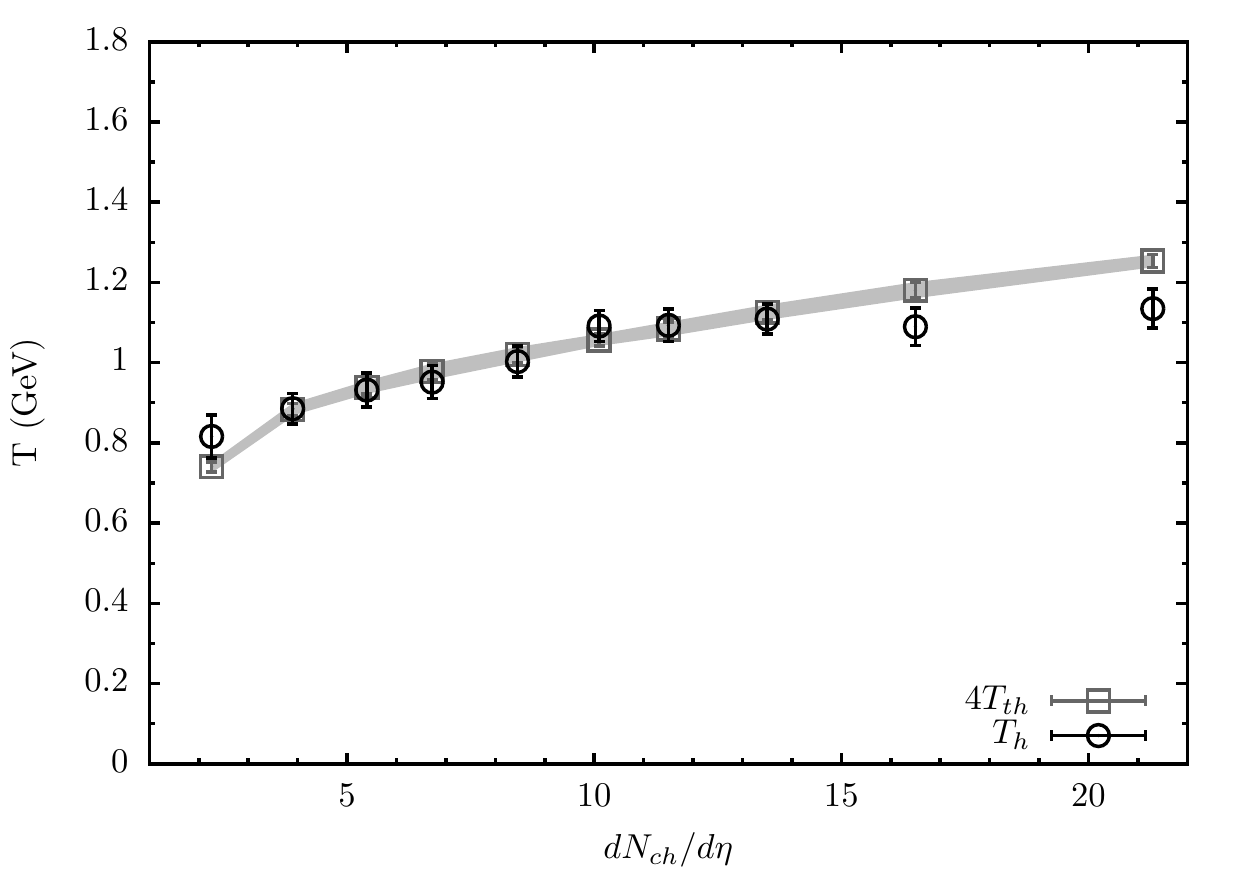}
\caption{ $T_{\rm h}$ and $4 T_{\rm th}$ as a function of centrality for $K_S^0$
  production in p-p collisions at $\sqrt{s_{NN}}=$7 TeV. }
\label{fig:figure3}
\end{figure}

In order to study the dependence on the multiplicity of $T_{\rm th}$ and
$T_{\ rm h}$ we have used the transverse momentum distribution of $K_S^0$
produced in p-p collisions at $\sqrt{s}=7 $TeV in the range up to $p_t\le 10$
GeV/c \cite{alice2015}. We use $K_S^0$ instead of $\pi$ or charged particles
because we have not found published data covering a broad range of soft and
hard regions at different multiplicities. In Figures (\ref{fig:figure1}),
(\ref{fig:figure2}) and (\ref{fig:figure3}), we show the fit and the results
for $T_{\rm h}$ and $T_{\rm th}$ respectively as a function of $dN_{\rm
  ch}/d\eta$. We observe an increase of $T_{\rm h}$ and $T_{\rm th}$. The
values of $T_{\rm th}$ and $T_{\rm h}$ are in the range [0.18,0.28] and
[0.8,1.15]. In Figure (\ref{fig:figure3}) the scaled curve $4T_{\rm th}$ is
shown compared to $T_{\rm h}$. We observe that the scaled $4T_{\rm th}$ curve
lies on the obtained $T_{\rm h}$ values, therefore the relation
between $T_{\rm th}$ and $T_{\rm h}$ (\ref{thermalVshard}) remains
approximately valid not only for different energies but also for different
centralities, pointing out to some physical reason behind. The obtained values
for $T_{\rm th}$ and $T_{\rm h}$ are slightly higher than the values $T_{\rm
  th}=0.17$ GeV and $T_{\rm h}=0.74$ GeV of reference \cite{baker2017} due to
the different sets of data, since in this analysis $K_S^0$ are used instead of
charged particles.

\begin{figure}[h]
\includegraphics[scale=1]{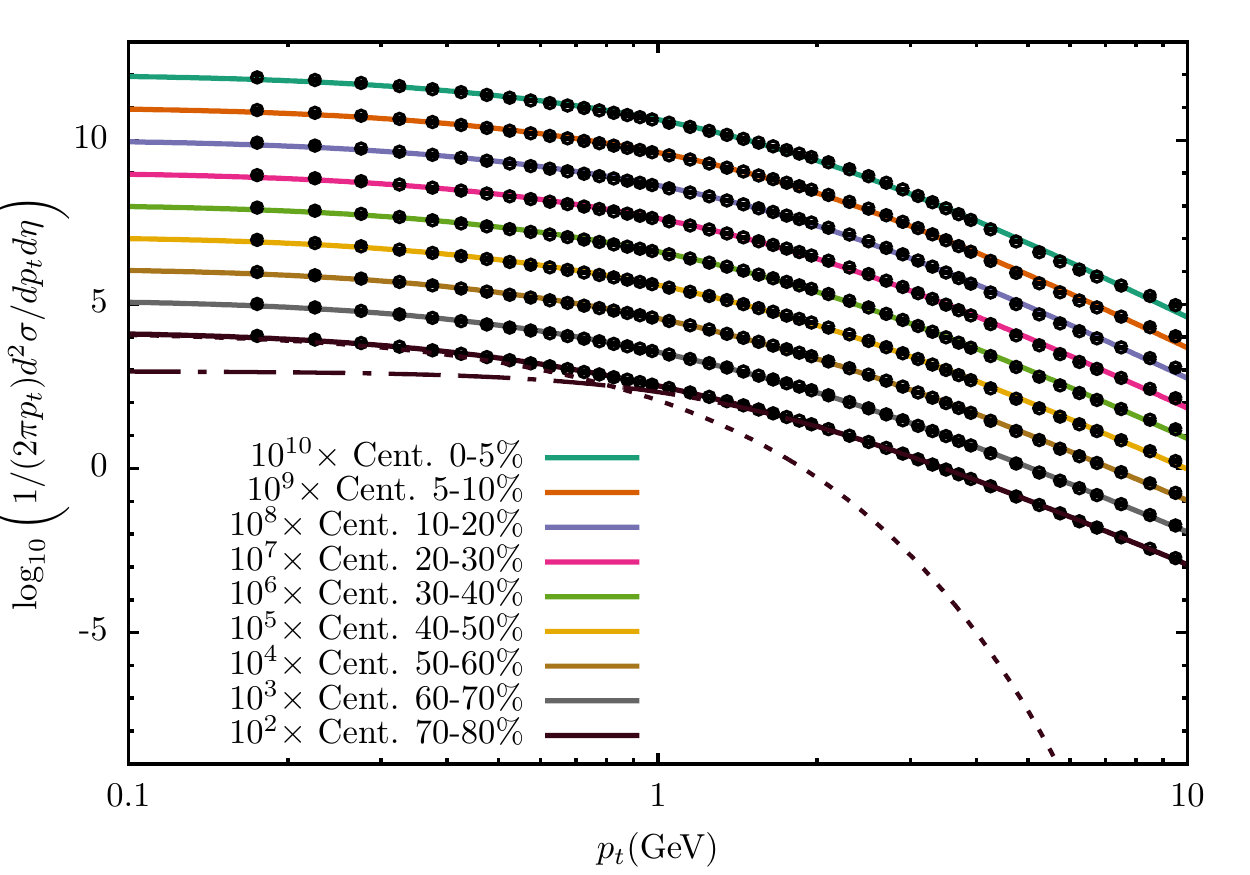}
\caption{Normalized differential charged particle production in Pb-Pb
  collisions at $\sqrt{s_{NN}}$=2.76 TeV, as a function of transverse momentum
  for different classes of centralities. Centrality classes from 0$\%$-5$\%$
  to 70$\%$-80$\%$, in decreasing magnitude, correspond to charged particle
  productions of $dN_{\rm ch}/d\eta$ = 1600, 1290, 960, 650, 425, 260, 145, 75
  and 30. Also shown are the thermal and hard components of the lowest
  centrality fit, by short dashed and long dashed lines, respectively.}
\label{fig:figure4}
\end{figure}

\begin{figure}[h]
\includegraphics[scale=1]{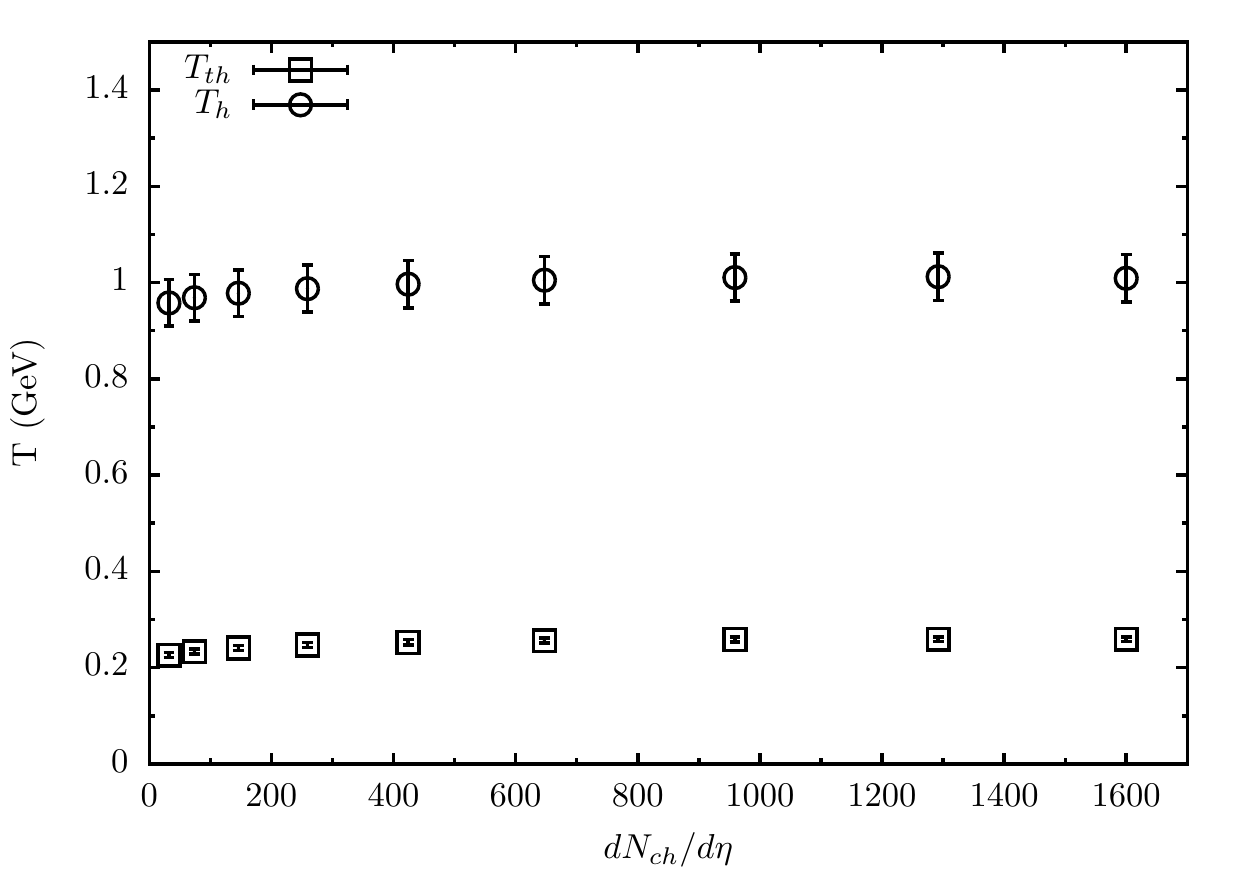}
\caption{Variation of $T_{\rm th}$ and $T_{\rm h}$ with centrality for
  charged particle production in Pb-Pb collisions at $\sqrt{s}=$ 2.76 TeV. }
\label{fig:figure5}
\end{figure}

\begin{figure}[h]
\includegraphics[scale=1]{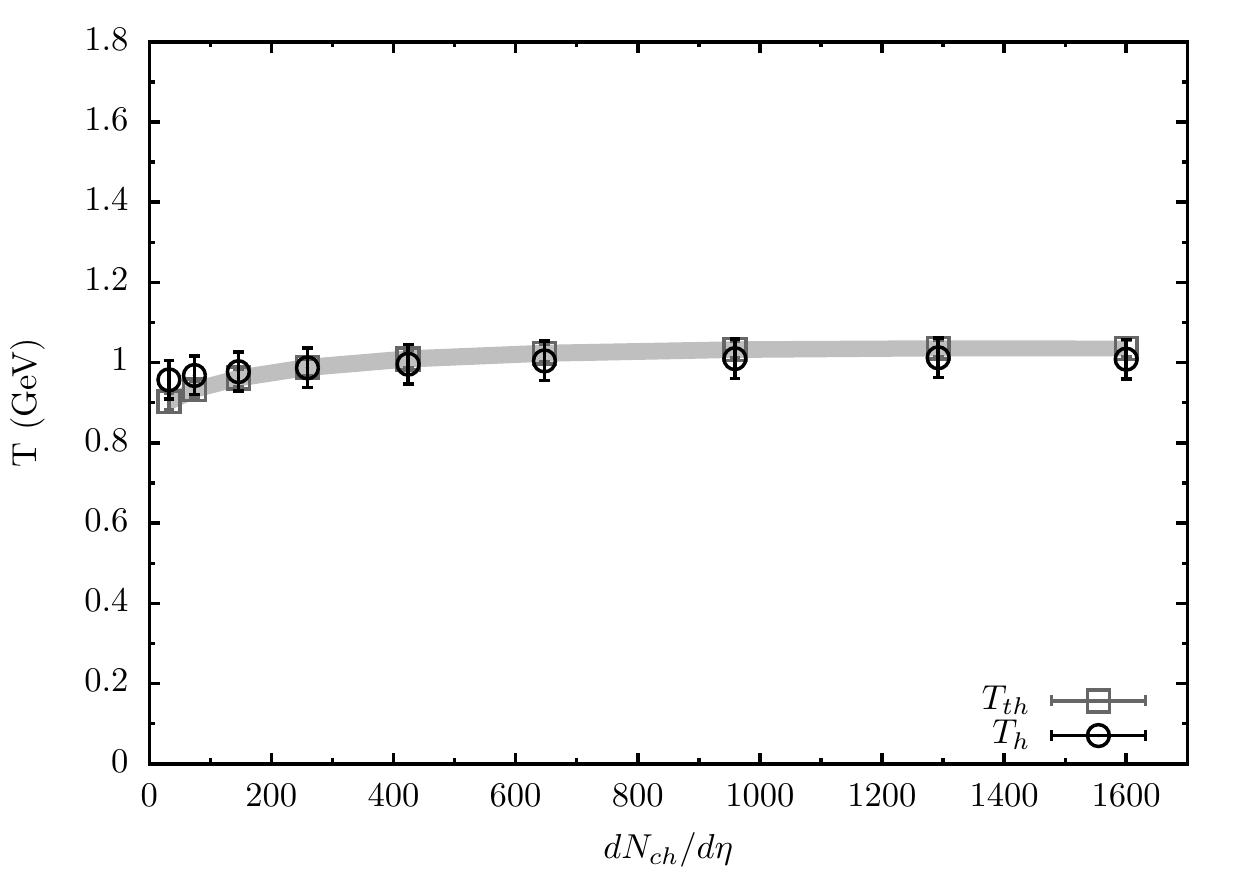}
\caption{ $T_{\rm h}$ and $4 T_{\rm th}$ as a function of centrality, for
  charged particle production in Pb-Pb collisions at $\sqrt{s}=$ 2.76 TeV. }
\label{fig:figure6}
\end{figure}

\begin{figure}[h]
\includegraphics[scale=1]{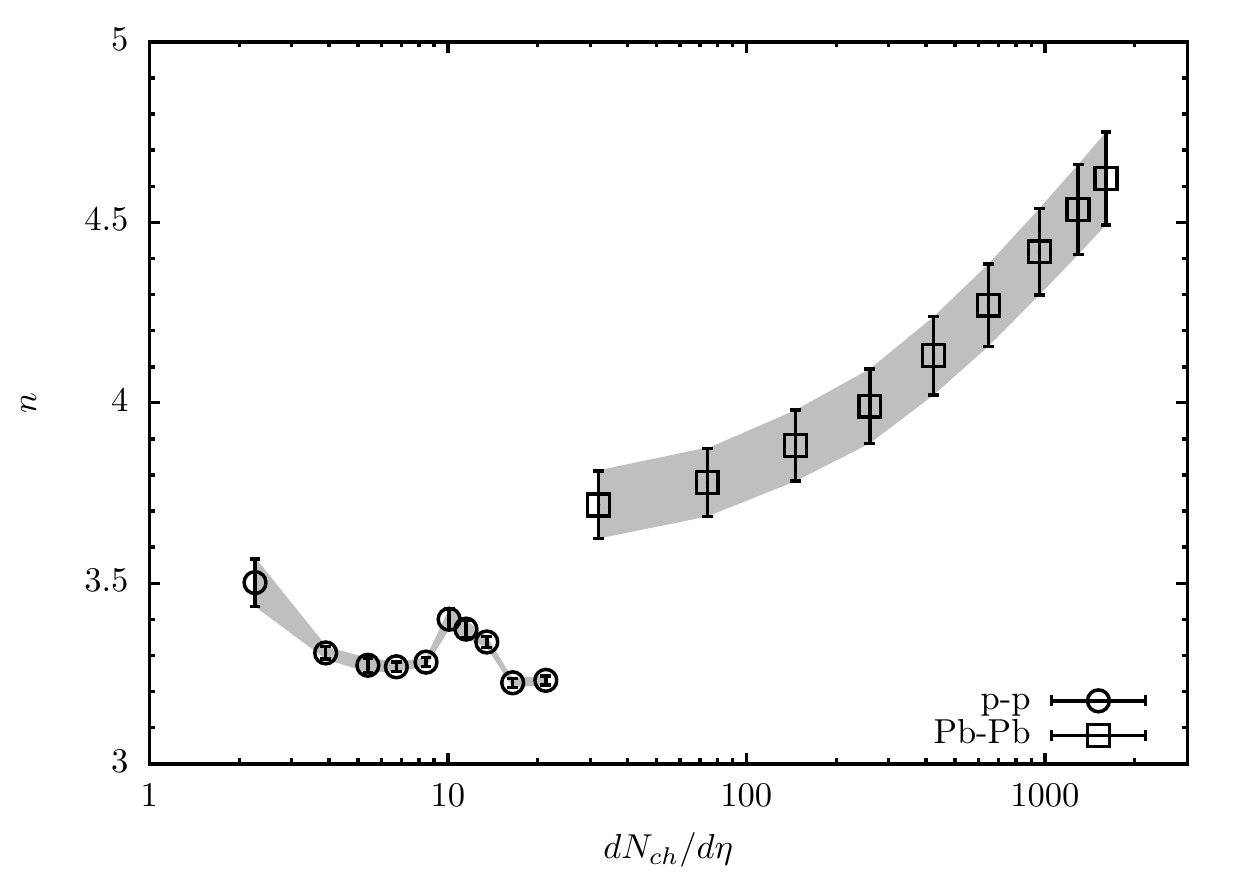}
\caption{ Power index $n$ of the hard component as a function of $dN_{\rm
    ch}/d\eta $ for $K_S^0$ production in p-p at $\sqrt{s}=$ 7 TeV, and
  charged particle production in Pb-Pb collisions at $\sqrt{s_{NN}}$=2.76
  TeV.}
\label{fig:figure7}
\end{figure}

We have extended the study to Pb-Pb collisions at different multiplicities by
fitting the ALICE collaboration TMD data for charged particles
\cite{alice2013} at $\sqrt{s}=2.76$ TeV. In Figures (\ref{fig:figure4}),
(\ref{fig:figure5}) and (\ref{fig:figure6}), we show the fit and the values
obtained for $T_{\rm th}$ and $T_{\rm h}$ as a function of the
multiplicity. $T_{\rm th}$ also increases with multiplicity and
$T_{\rm h}$ follows the same relation $T_{\rm h}\approx 4T_{\rm th}$ observed
at p-p collisions. 
In Figure (\ref{fig:figure7}) we show the results of the fit for the power index
$n$ in p-p and Pb-Pb collisions as a function of the charged particle
production. $n$ decreases with multiplicity for p-p collisions and, on the
contrary, increases for Pb-Pb collisions. For p-p collisions at $\sqrt{s}=$13
TeV the value obtained for the case of charged particles is $n=$3.1, slightly
smaller than the value of Figure (\ref{fig:figure7}).
The values of $n$ are larger for Pb-Pb than for p-p collisions as expected due
to the jet quenching and correspondingly to high $p_t$ particle suppression. 

\section{Discussion}
\label{discussion}
Our results for p-p collisions show that for each multiplicity the effective
thermal temperature obtained from the TMD of the produced particles can be
viewed as a rapid quench of the entangled partonic state. The behavior of
$T_{\rm th}$ and $T_{\rm h}$ as a function of the multiplicity is very similar,
holding the relation $T_{\rm h}\approx 4T_{\rm th}$ for all the studied
multiplicities. This fact adds evidence to the cases studied in reference
\cite{baker2017}. It is remarkable that the same relation holds for Pb-Pb
collisions, realizing the different values of $T_{\rm th}$ and $T_{\rm h}$
compared with the p-p case. The increase of $T_{\rm th}$ with multiplicity in
p-p collisions is larger than in the Pb-Pb case as it was expected, as far as
$T_{\rm th}$ is nothing but $\langle p_t \rangle$ and experimentally the LHC
data \cite{alice2013} has shown a larger increase with multiplicity in p-p
than in Pb-Pb collisions.

Equations (\ref{thermal_parametrization}) and (\ref{hard_parametrization}) can
be obtained in the framework of clustering of color sources
\cite{braun2015,armesto1996,nardi1998} as we show in section \ref{clustering}.
In this approach, $T_{\rm th} \approx T_{\rm h}/\pi \sqrt 2$ and the cluster
size distribution is a gamma function \cite{diasdedeus2005,diasdedeus2004,braun2000a,braun2000b} which coincides with the stationary
solution of the Fokker-Planck equation derived from the Langevin equation
corresponding to a white noise due to the quench of a hard parton
collision \cite{Wilk,Biro}. In this way the fluctuation in the cluster size are related to the
time temperature fluctuations. 

\begin{figure}[h]
\includegraphics[scale=1]{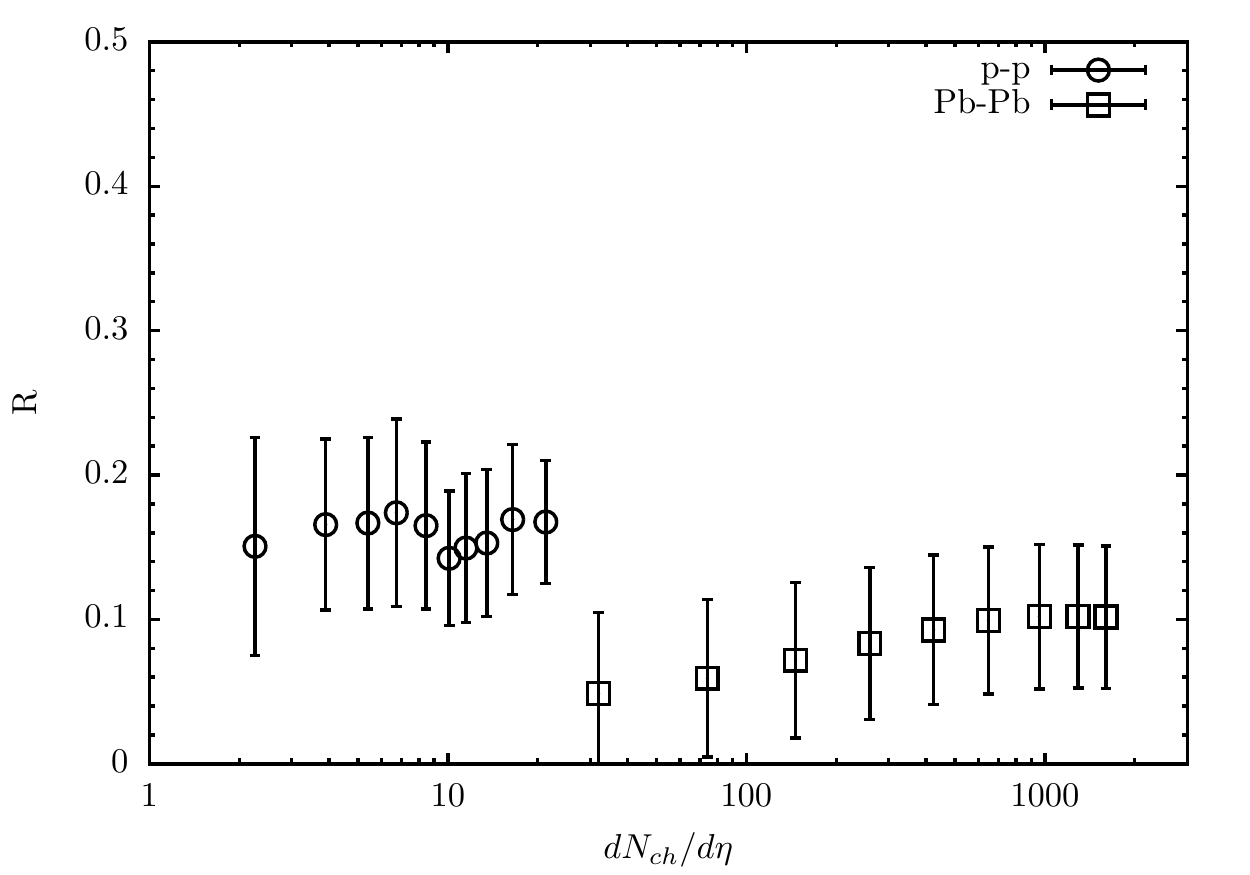}
\caption{ $R$, the ratio of the hard component to the total component as a
  function of $dN_{\rm ch}/d\eta$, for $K_S^0$ production in p-p at
  $\sqrt{s}=$7 TeV and for charged particle production in Pb-Pb collisions at
  $\sqrt{s_{NN}}$=2.76 TeV.}
\label{fig:figure8}
\end{figure}

The ratio $R$ of the integral under the power law curve (hard component) and
the integral over the total (hard + thermal components)
\begin{equation}
R=\frac{H}{H+S},
\end{equation}
is plotted in Figure (\ref{fig:figure8}), for different multiplicities for p-p
and Pb-Pb collisions. The value found at $\sqrt{s}=13$ TeV in reference
\cite{baker2017} was $R\approx$ 0.16, in agreement with the ratio calculated
in inelastic proton-proton collisions at $\sqrt{s}$=23, 31, 45, and 53
GeV. Our results for p-p collisions at $\sqrt{s}$=7 TeV for different
multiplicities are close to the mentioned value. In the case of Pb-Pb we have
found a smaller value. This smallness is consistent with the behavior found
in \cite{Carlota} based on saturation momentum and geometrical scaling.

\section{Clustering of color sources}
\label{clustering}

Multiparticle production is currently described in terms of color sources
(strings) stretched between the projectile and the target. These strings decay
by the Schwinger $q$-$\bar q$ production and subsequently hadronize to produce
the observed hadrons. The color in the strings is confined in a small area in
the transverse space of the order of $0.2$ fm. With increasing energy and/or
atomic number of the colliding objects, the number of color sources grows, and
they start to overlap, forming clusters. A cluster of $n$ color sources
behaves as a single string with an energy-momentum that corresponds to the sum
of the energy-momentum of the overlapping strings and with a higher color
field, corresponding to the vectorial sum of the color charges of each
individual string. The resulting color field covers the area $S_n$ of the
cluster. Thus $\vec Q_n^2 = (\sum_1^n \vec Q_i)^2$, and given that the
individual string colors can be arbitrarily oriented in color space, the
average $\vec Q_i \cdot \vec Q_j$ is zero, so $\vec Q_n^2 =n \vec Q_1^2$. As
$Q_n$ depends also on the area, we have $Q_n = \sqrt{n S_n/S_1} Q_1$, where
$S_1 $ is the area of the individual string. The mean multiplicity and the
mean transverse momentum are proportional to the color charge and to the color
field respectively
\begin{equation}
\mu_n = \sqrt{\frac{n S_n}{S_1}} \mu_1 \;\;\;\;\;\;
\average{p_t^2}_n = \sqrt{\frac{n S_1}{S_n}} \average{p_t^2}_1,
\end{equation}
which in the limit of high density, $\xi = N_s S_1/S$, becomes
\begin{equation}
\mu_n = N_s F(\xi) \mu_1 \;\;\;\;\;\;
\average{p_t^2}_n = \frac{1}{F(\xi)} \average{p_t^2}_1,
\end{equation}
where $N_s$ is the number of color sources and $F(\xi)$ is an universal factor
\begin{equation}
F(\xi) = \sqrt{\frac{1-e^{-\xi}}{\xi}}.
\end{equation}
The factor $1-e^{-\xi}$ is the fraction of the total collision area covered by
color sources at density $\xi$ (it is assumed an homogeneous profile for the
collision area). $F(\xi)$ goes to $1$ at low densities and goes to $0$ at
high $\xi$. The transverse momentum distribution $f(p_t)$ is obtained from the
Schwinger's distribution, $\exp(-p_t^2 x$), weighted by the cluster size
distribution $W(x)$, where $x$ is the inverse of $\average{p_t^2}_n$
\begin{equation}
f(p_t) = \int dx W(x) \exp(-p_t^2 x).
\label{ptdef}
\end{equation}
The weight function is the gamma function because the process of increasing
the centrality or energy of the collision can be regarded as a transformation
of the color field located in the sites of the surface area, implying a
transformation of the cluster size distribution of the type
\begin{equation}
W(x') \rightarrow \frac{x' W(x')}{\average{x'}} \rightarrow \cdots \frac{x'^k
  W(x')}{\average{x'^k}} \rightarrow \cdots
\end{equation}
This renormalization group type of transformation have been studied long time
ago in probability theory showing that the only stable distributions under
such transformations are the generalized gamma function. We take the simplest
case, namely, the gamma function
\begin{equation}
W(x) = \frac{\gamma}{\Gamma(n)} (\gamma x)^n \exp(-\gamma x),
\label{weight}
\end{equation}
with 
\begin{equation}
\gamma = \frac{n}{\average{x}},
\end{equation}
and 
\begin{equation}
\frac{1}{n} = \frac{\average{x^2}-\average{x}^2}{\average{x}^2}.
\end{equation}
Introducing eq. \ref{weight} into equation \ref{ptdef} we obtain the
distribution \cite{diasdedeus2005}
\begin{equation}
f(p_t) = \frac{1}{(1+p_t^2/\gamma)^n} =  \frac{1}{(1+\frac{F(\xi) p_t^2}{n \average{p_t^2}_1})^n},
\label{pt_distri_hard}
\end{equation}
which takes the form of the parameterization used in the
eq. \ref{hard_parametrization} with
\begin{equation}
T_{\rm h}^2 = \frac{\average{p_t^2}_1}{F(\xi)},
\end{equation}
which grows with the density $\xi$ and thus with the energy and centrality as
it is observed in the analysis of p-p and Pb-Pb collisions. In the last case,
if the fits include larger $p_t$ values it is observed a flattening of the
dependence of $T_{\rm h}$ with multiplicity due to jet quenching effects. At
low $p_t$, equation \ref{pt_distri_hard} behaves as 
\begin{equation}
f(p_t) \approx \exp(-p_t^2 F(\xi)/\average{p_t^2}_1),
\label{pt_distri_soft}
\end{equation}
independently of $n$. In this low $p_t$ regime, there are other effects, like
fluctuations of the color field, which should be taken into account. In fact,
assuming that such fluctuations are Gaussian, we have \cite{diasdedeus2016,castorina2007}
\begin{equation}
\sqrt{\frac{2}{\pi \average{x_h^2}}} \int_0^\infty \exp(-\frac{x_h^2}{2
  \average{x_h^2}}) \exp(-\frac{\pi p_t^2}{2 x_h^2})  =
\exp(-p_t\sqrt{\frac{2 \pi}{\average{x_h^2}}})
\label{pt_gauss_convol}
\end{equation}
where we defined $x_h^2 = \pi T_{\rm h}^2$. The equation \ref{pt_gauss_convol}
expresses the thermal behavior with 
\begin{equation}
T_{\rm th} = \frac{T_{\rm h}}{\pi \sqrt 2}.
\end{equation}
In other words, the thermal temperature is given by the fluctuations of the
hard temperature which are proportional to this hard temperature. 

We notice that according to equation \ref{pt_distri_hard} the power index $n$
is related to the inverse of the width of the distribution and a different
behavior with multiplicity is obtained for p-p and Pb-Pb collisions. This
fact is a consequence of the clustering of the color sources. At low density
of sources, there are only a few clusters of overlapping strings and thus the
only temperature fluctuations come from inside the individual strings. As the
number of clusters with different number of color sources increases, the
fluctuations also increase and correspondingly $n$ decreases. If the color
density increases further, the clusters of different color start to overlap in
such a way that the number of clusters with different number of color sources
decreases and thus the fluctuations decrease and $n$ increases. The change of
behavior which can be observed in fig. \ref{fig:figure7} is related to the
critical percolation point. Notice that $n$ is decreasing with multiplicity in
p-p collisions but we expect that above a given multiplicity $n$ should start
to grow in the same way as in Pb-Pb collisions. 

\section{Time evolution}
\label{sec:langevin}

The scale of the TMD is $u=T_{\rm h}^2$ which obeys the Langevin equation 
\begin{equation}
\frac{du}{dt} + \left( \frac{1}{\tau}+ \zeta(t) \right) u = \phi,
\label{langevin}
\end{equation}
where $\tau$ is a characteristic damping time and
$\zeta(t)$ is a white Gaussian noise, with mean $\average{\zeta(t)} = 0$
and a correlator corresponding to fast changes such as expected in the quench
induced by a  hard collision, 
\begin{equation}
\langle \zeta(t) \zeta(t+\Delta t) \rangle = 2 D \delta(\Delta t),
\end{equation}
where $D$ is the variance and $\phi$ a constant.

The associated Fokker Planck equation for the
probability to have the temperature $T$ at time $t$ under the above noise is
given by \cite{Wilk,Biro}
\begin{equation}
\frac{\partial f(T,t)}{\partial t} = -\frac{\partial }{\partial T} K_1 f(T,t)
+ \frac{1}{2} \frac{\partial^2}{\partial T^2} K_2 f(T,t),
\label{fokker}
\end{equation}
being
$$
K_1(T) = \phi -2 \frac{T^2}{\tau} + D T^2, \; \; \; \; \; \; 
K_2(T) = 2 D T^4.
$$
The stationary solution of equations (\ref{langevin},\ref{fokker}) is just
the gamma distribution on the variable $1/T^2$
\begin{equation}
f(T) = \frac{\mu}{\Gamma(n)} \left( \frac{\mu}{T^2}\right)^n \exp(-\mu/T^2),
\end{equation}
with 
\begin{equation}
\mu = \phi/D, \; \; \; \; \; \; 
n = 1/\tau D.
\end{equation}
This distribution coincides with the cluster size distribution of equation
\ref{weight}. 
In this way, the temperature fluctuations given by the inverse of the index
$n$ are related to the product $\tau D$ which have to do with the time
evolution.  

\section{Conclusions}
\label{conclusions}
The analysis of the dependence on the multiplicity of the LHC p-p and Pb-Pb
data confirms the picture of thermalization induced by quantum
entanglement. In all the analyzed data, the effective thermalization
temperature obtained from data is proportional to the hard scale of the
collision $T$ given by the average momentum transfer. The coefficient of
proportionality is universal, independent of the considered collision, even
though $T_{\rm th}$ and $T_{\rm h}$ are different in each collision
type. Thermal and hard temperatures increase with multiplicity in both
collision scenarios, and this rise reproduces the known correlation of
$\langle p_t \rangle$ and $dN_{ch}/d\eta$ for p-p and Pb-Pb collisions. In the
framework of clustering of color sources the proportionality between $T_{\rm
  th}$ and $T_{\rm h}$ is understood, being $T_{\rm th}/T_{\rm h} = 
\pi\sqrt{2}$.  The $n$ parameter of the hard distribution decreases with
multiplicity for $p-p$ collisions and increases for Pb-Pb collisions. This
fact means that the normalized transverse momentum fluctuations behave quite
different with multiplicity in p-p and Pb-Pb collisions. This behavior is
naturally explained by the clustering of color sources. The change in the
behavior of $n$ is related to the formation of a large cluster of the initial
color sources (partons), which marks the percolation phase transition.

The cluster size distribution is a gamma function which is also the stationary
solution of the Fokker-Planck equation associated to the Langevin equation 
for a white Gaussian noise due to the quench of a hard partonic collision. 

\section{Acknowledgments}
 We thank the grant Mar\'{\i}a de Maeztu Unit of Excelence of Spain and the
 support of Xunta de Galicia under the project ED431C2017. This paper has been
 partially done under the project FPA2014-58293-C2-1-P of MINECO (Spain).

\end{document}